\documentclass[%
 reprint,
 amsmath,amssymb,
 aps]{revtex4-2}
\usepackage{tabularx} 
\usepackage[normalem]{ulem}
\usepackage{makecell} 
\usepackage{dcolumn}
\usepackage{xcolor}
\usepackage{bm}
\usepackage{multirow} 
\usepackage{graphicx}
\usepackage{subcaption} 
\usepackage{booktabs} 
\usepackage{amsmath}
\usepackage{amssymb}
\newcommand{\aap}{Astronomy \& Astrophysics}

\begin{document}

\preprint{APS/123-QED}

\title{Time-dependent multi-energy neutrino emission from symbiotic recurrent novae: the role of accretion disks}

\author{Rui Xu}
\thanks{These authors contributed equally to this work.}
\affiliation{School of Physics and Astronomy, Sun Yat-sen University, Zhuhai, 519082, China}

\author{Yudong Cui}%
\affiliation{School of Physics and Astronomy, Sun Yat-sen University, Zhuhai, 519082, China}

\author{Yihan Shi\thanks{These authors contributed equally to this work.}}
\thanks{These authors contributed equally to this work.}
\affiliation{School of Physics and Astronomy, Sun Yat-sen University, Zhuhai, 519082, China}

\author{Lili Yang}
\email{yanglli5@mail.sysu.edu.cn}
\affiliation{School of Physics and Astronomy, Sun Yat-sen University, Zhuhai, 519082, China}
\affiliation{Centre for Astro-Particle Physics, University of Johannesburg, PO Box 524, Auckland Park 2006, South Africa}

\date{\today}

\begin{abstract}
Symbiotic recurrent novae provide a unique laboratory for studying thermonuclear explosions, shock evolution, and nonthermal particle acceleration in dense circumstellar environments. In this work, we develop a time-dependent, multi-energy framework to describe neutrino emission from such systems, consistently incorporating both MeV neutrinos produced during thermonuclear runaway and GeV neutrinos generated through hadronic interactions in nova-driven shocks.
Using RS Oph as a benchmark source, we model the evolution of the shock interacting with both the red giant wind and a dense accretion disk surrounding the white dwarf. We show that the resulting neutrino signal exhibits a characteristic two-component temporal structure: an early, rapidly rising MeV component tracing nuclear burning, followed by a delayed GeV component governed by shock propagation and particle acceleration.
The presence of an accretion disk can significantly enhance the early-time GeV neutrino emission by providing a dense target for proton–proton interactions. This leads to a pronounced neutrino flux within the first few hours after eruption, a feature absent in wind-dominated scenarios. We further evaluate the detectability of these signals and find that while the MeV component remains below current detection thresholds, the GeV neutrino emission from nearby systems may become accessible to next-generation detectors.
Our results highlight the critical role of the circumstellar structure in shaping nova neutrino emission and demonstrate that symbiotic recurrent novae are promising targets for future multi-messenger observations.

\end{abstract}

\maketitle

\section{Introduction}

A nova outburst is a thermonuclear explosion that occurs on the surface of an accreting white dwarf (WD) in a close binary system. 
The hydrogen-rich material transferred from the companion star accumulates on the WD surface and is gradually compressed until the temperature and density become high enough to trigger a thermonuclear runaway (TNR) \cite{Guetta:2022ims}. Depending on the nature of the companion star, novae are commonly classified into classical novae and symbiotic novae \cite{Munari:2024gmf}.

Among these systems, Symbiotic Recurrent Novae (SymRNe) are of particular interest. They consist of a massive WD and a red giant (RG) companion, and are characterized by recurrent eruptions on timescales of years to decades \cite{schaefer2025comprehensive}. Because the RG supplies a dense stellar wind of about $10^8 - 10^9~\mathrm{cm^{-3}}$ and the WD typically accretes at a relatively high rate, the interaction between the nova ejecta and the surrounding medium is especially strong. As a result, SymRNe provide favorable conditions for studying shock evolution, high-energy radiation, and particle acceleration. Representative Galactic sources include RS Oph, T CrB, V407 Cyg, and V745 Sco, as listed in Table \ref{Table 2}. They all have erupted multiple times in the history, and their central WDs have relatively large masses ($1.22 M_\odot - 1.39 M_\odot$), close to the Chandrasekhar limit $\sim 1.44 M_\odot$. Near this limit, the radius of a WD decreases sharply according to theoretical models, so the WD radii of these novae are approximately $\sim 0.01 R_\odot$ \cite{Astashenok:2022uhw, parsons2017testing}. 

\begin{table*}[htbp]
\begin{tabular}{ccccc}
\hline
SymRNe    & RS Oph      & T CrB       & V407 Cyg     & V745 Sco     \\ \hline
Outbursts    & \begin{tabular}[c]{@{}c@{}}1898 1933 1958\\ 1967 1985 2006 2021\end{tabular} & 1866 1946      & 1936 2010      & 1897 1937 1989 2014     \\D{[}kpc{]}       & 2.71      & 0.91    & 3.10   & 8.02     \\Orbital Period    & 455       & 228       & 775     & 930   \\
WD Mass ($M_\odot$)     & 1.37    & 1.38          & 1.22        & 1.39      \\
\begin{tabular}[c]{@{}c@{}}Position\\ (J2000 Equatorial Coordinates)\end{tabular} & \begin{tabular}[c]{@{}c@{}}RA: 17h 50m 13s\\ Dec: -06°42'28"\end{tabular}    & \begin{tabular}[c]{@{}c@{}}RA:15h 59m 30s\\ Dec: +25°55'12"\end{tabular} & \begin{tabular}[c]{@{}c@{}}RA: 21h 02m 09s\\ Dec: +45°46'32"\end{tabular} & \begin{tabular}[c]{@{}c@{}}RA:17h 55m 22s\\ Dec: -33°14'58"\end{tabular} \\
\hline
\end{tabular} 
\caption{The outburst year, distance, orbital period, mass of the compact engine\cite{schaefer2025comprehensive,Munari:2010ed}, and position\cite{Munari:2024gmf} of four SymRNe in our Galaxy.}
    \label{Table 2}
\end{table*}

Neutrino production in novae can arise from different physical channels. Before and during the early stages of the outburst, low-energy MeV neutrinos are expected to be generated by nuclear and thermal processes associated with hydrogen burning on the white dwarf surface, including reactions in the $pp$ chain and the CNO cycle, as well as thermal neutrino processes such as pair annihilation, plasma decay, and bremsstrahlung. These neutrinos can probe the thermonuclear conditions directly.

After eruption, the expanding ejecta with a large number of particles compressed into a thin shell, resulting in high density and rapid expansion along with the shock \cite{MAGIC:2022rmr}, driving shocks into the surrounding medium. If protons are efficiently accelerated on the shock front \cite{DeSarkar:2023nhp}, hadronic interactions with ambient matter can produce neutral and charged pions. The decay of charged pions then leads to high-energy neutrino emission, while neutral pion decay produces gamma rays. In dense symbiotic environments, proton-proton interactions are generally expected to be the dominant hadronic channel \cite{Bednarek:2022vey, HESS:2022qap}, making SymRNe promising candidates for neutrino production.

Recently, several novae have been detected in GeV gamma rays by \textit{Fermi}-LAT, beginning with V407 Cyg \cite{Abdo:2010he}, and the 2021 eruption of RS Oph was further observed by both \textit{Fermi}-LAT and H.E.S.S. \cite{HESS:2022qap}. These detections indicate that efficient particle acceleration can operate in nova shocks and motivate the search for accompanying neutrino emission. 
{\color{black} While recent studies on wind-only models and magnetic reconnection scenarios have primarily focused on TeV-scale observatories \cite{Partenheimer:2024qxw,IceCube:2025egb,sarmah2025multi}, the unique circumstellar structure—particularly the dense accretion disk surrounding the white dwarf—could significantly enhance the early GeV emission.}{\color{black}Furthermore, existing studies have mostly treated the MeV and GeV components separately, focusing on either the thermonuclear burning phase or the shock propagation. A unified time-dependent framework that consistently incorporates both components, along with their associated electromagnetic radiation, remains lacking \cite{Wang:2025ibu, Guetta:2022ims}.}

Another important issue concerns the structure of the circumstellar environment. In SymRNe, the WD is not only embedded in the RG wind, but may also be surrounded by an accretion disk. Such a disk introduces a dense and anisotropic target for hadronic interactions and may therefore significantly affect the early high-energy neutrino output. The contribution of the accretion disk to the temporal evolution of nova neutrino emission may play an essential role.

Moreover, current neutrino telescopes, limited by the irreducible background of atmospheric neutrinos, primarily focus on astrophysical sources in the TeV-PeV energy range \cite{IceCube, KM3NeT}. Consequently, both theoretical investigations and detection strategies for GeV-scale astrophysical neutrinos remain relatively underexplored. However, transient sources such as SymRNe, whose high-energy emission is predominantly concentrated in the GeV band, present a compelling opportunity. Thanks to their short duration, these events may become detectable if they occur sufficiently close to Earth. Successful observation would significantly enhance our understanding of astrophysical sources and their surrounding environments. Therefore, we evaluate the detection prospects of such sources with ORCA and further propose that an experiment with an effective volume approximately two orders of magnitude larger than ORCA would enable their observation.

In this work, we develop a time-dependent multi-energy framework for neutrino emission from SymRNe, taking RS Oph as a benchmark source. We model the MeV neutrino component associated with TNR-driven nuclear processes and the GeV component produced through hadronic interactions during shock propagation. In particular, we include both the RG wind and the accretion disk in the target environment, and examine how these components shape the resulting neutrino light curves and spectra.

Our results indicate that the neutrino signal shows characteristic two-stage behavior. The MeV component rises rapidly and traces the evolution of thermonuclear burning, whereas the GeV component emerges later and is governed by the interaction of the shock with the surrounding medium. We find that the accretion disk can considerably enhance the early GeV neutrino emission by increasing the target density for proton-proton collisions, even though the gamma-ray counterpart produced will be absorbed in the dense environment. We also discuss the detectability of these signals and present predictions for the forthcoming eruption of T CrB.

This paper is organized as follows. In Section ~\ref{Physical scenario and model setup}, we describe the physical scenario and model setup. In Section ~\ref{Application to RS Oph}, we apply the model to RS Oph and derive the time-dependent neutrino light curves. In Section~\ref{Detectability and implications}, we discuss the neutrino spectra and their detectability. In Section \ref{Discussion and limitations}, we summarize the main results and present our conclusions.

\section{Physical scenario and model setup } \label{Physical scenario and model setup}
\subsection{Symbiotic recurrent nova geometry}

\begin{figure}
    \centering
    \includegraphics[width=0.45\textwidth]{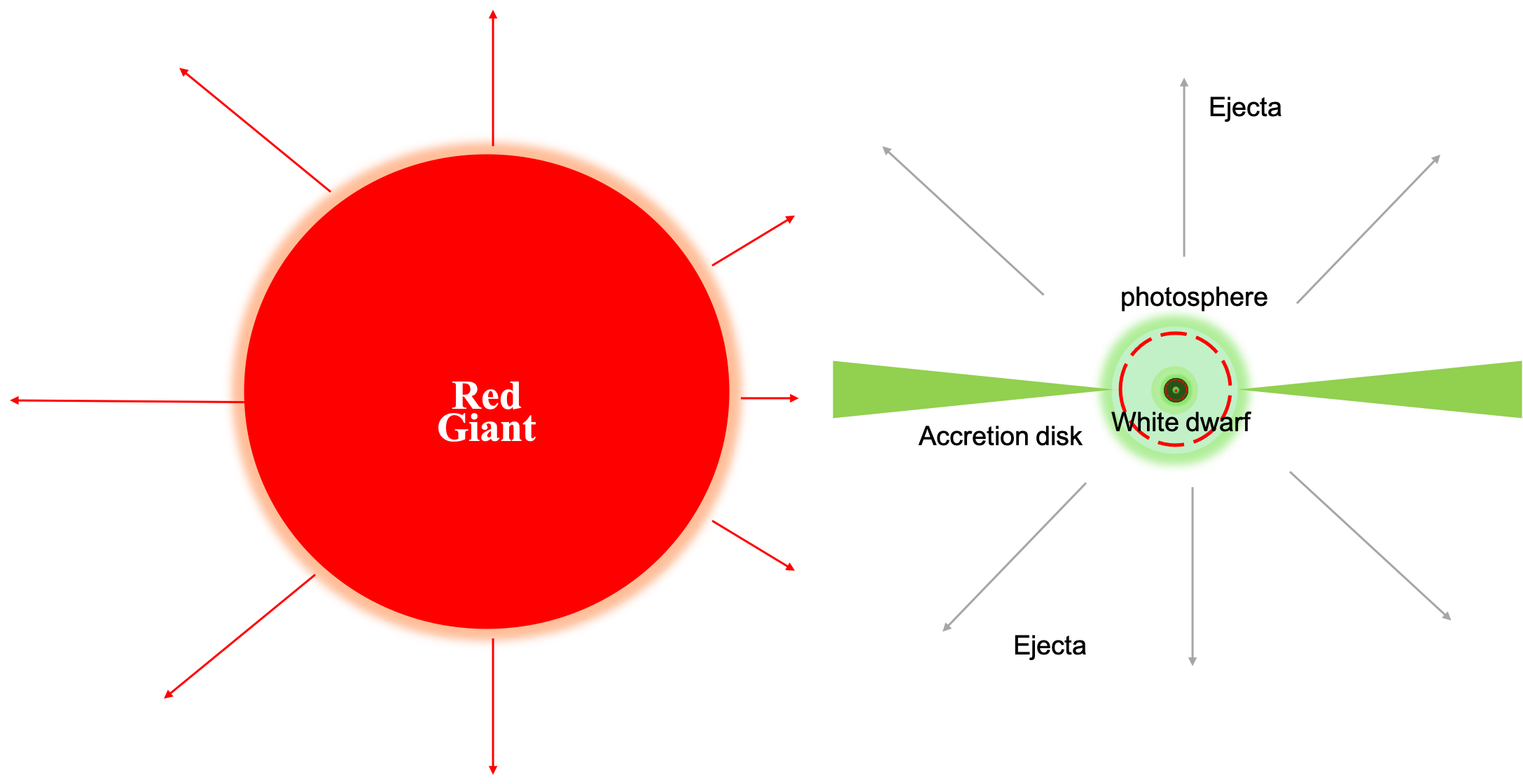}
    \caption{Schematic diagram of the geometric structure of a SymRN.}
    \label{fig:structure}
\end{figure}

As illustrated in Fig. \ref{fig:structure}, the system exhibits a highly structured geometry consisting of the central WD, the RG companion, a dense RG wind, and an accretion disk surrounding the WD. Material lost through the stellar wind of the RG is captured by the gravitational field of the WD and accumulates on its surface in the form of a hydrogen-rich envelope. When the temperature and density at the base of this accreted layer reach critical values, a TNR is triggered, leading to the ejection of the outer envelope at velocities of order $10^{3}$–$10^{4}~\mathrm{km~s^{-1}}$.
For the well studied RS Oph, its characteristic system parameters are summarized in Table \ref{Table 1}, including the WD radius, the RG radius, the inner and outer radii of the accretion disk, the maximum heights of the disk, the binary separation and the maximum photospheric radius. 

Because the RG atmosphere is extended and loosely bound, mass loss is expected to create a dense circumstellar environment around the WD. In addition to the stellar wind, we assume that part of the transferred material forms an accretion disk around the WD. The spacing between the photshpere of WD and the accretion disk is 2.2 $R_{\odot}$. According to Eq. \ref{eq:shock_radius}, 0.14 hours after the eruption, the shock will interact with the disk and enhance the GeV neutrino emission for 4.42 hours.

\begin{table}[]
\begin{tabular}{cc}
\hline
Parameter                                  & Value ($R_\odot$)         \\ \hline
WD's radius ($R_{WD}$)                      & 0.01 \\
RG's radius ($R_{RG}$)                      & 106  \\
Inner radius of accretion disk ($R_{in}$)  & 16.6   \\
Outer radius of accretion disk ($R_{out}$) & 190  \\
Maximum height of accretion disk (H) & 43.9\\
Distances between two objects (a)            & 332  \\
Maximum radius of photosphere ($R_{max}$)   & 14.4 \\ \hline
\end{tabular}\caption{Geometrical parameters of RS Oph\cite{Lechuga:2025jei}.} \label{Table 1}
\end{table}

\subsection{TNR-driven MeV neutrinos}
In the early accretion stage, the temperature and density at the base of the accreted layer remain relatively low, allowing hydrogen-rich matter to accumulate gradually on the WD surface. During this phase, hydrogen burning proceeds slowly through the $pp$ chain, and the released energy is approximately balanced by radiative and conductive losses. Low-energy neutrino production is correspondingly weak and is mainly associated with nuclear and thermal processes, including the $pp$ chain and plasma decay.

As accretion continues, compression of the envelope increases the temperature and density at its base. Once the temperature reaches approximately $10^{7} \mathrm{K}$, the CNO cycle becomes the dominant channel. Since the corresponding energy generation rate is strongly temperature dependent $ \varepsilon\varpropto T^{17}$, causing an exponential increase. Later, nuclear burning becomes unstable when radiative and conductive cooling can no longer compensate for energy release. Convection then develops rapidly, mixing fresh hydrogen into the burning region and accelerating the rise in temperature. This marks the onset of the TNR and the beginning of the nova outburst phase.

The nova explosion is initiated in a thin layer near the WD surface. Although thermal neutrinos can in principle be emitted from the underlying WD, their contribution is expected to be subdominant \cite{Wang:2025ibu}. Following the assumptions in the work \cite{Alsabti:2017ahu}, the accreted layer is dense and thin, with a characteristic density of order 40 $\mathrm{g cm^{-3}}$, so that the MeV neutrino signal directly traces the thermonuclear evolution in the envelope. 

\subsection{Shock–disk and shock–wind interactions }


Prior to the outburst, the matter that will later form the ejecta is confined to the accreted layer on the WD surface and has essentially no radial velocity. Once the TNR is triggered, the envelope is rapidly accelerated by the resulting pressure-driven outflow, reaching velocities from several hundred to several thousand kilometers per second on timescales from seconds to hours. The expanding ejecta then drive shocks into the surrounding circumstellar medium.

Nova shocks may in general include both internal and external components. {\color{black}Initially, the shock expands outward with minimal deceleration at a nearly constant high speed due to the low ambient density.} {\color{black}Internal shocks arise when later fast winds overtake earlier slower ejecta, while the external forward shock is generated by the interaction with ejecta and circumstellar medium.} {\color{black}For symbiotic recurrent novae, the dense circumstellar medium provided by the red giant wind makes the external forward shock particularly prominent. Extensive observational campaigns on the 2021 eruption of RS Oph have confirmed its dominant role in particle acceleration and high-energy emission \cite{MAGIC:2022rmr,HESS:2022qap}. Consequently, the external forward shock serves as the primary site for hadronic interactions and the production of high-energy neutrinos and gamma rays.}

{\color{black}
The ejecta, launched at speeds exceeding \(\sim 10^3\) km/s due to the TNR, sweeping through the RG wind and accretion disk \cite{Corso:2023cmt}. The acceleration timescale for a proton of energy \(E\) can be estimated as \cite{MAGIC:2022rmr},

\begin{equation}
t_{\text{acc}} = \frac{E}{(\mathrm{d}E/\mathrm{d}t)_{\text{acc}}} = 3.9 \left( \frac{E}{300\ \mathrm{GeV}} \right) \left( \frac{\xi B}{10^{-7}\ \mathrm{G}} \right)^{-1} \ \mathrm{day},
\label{eq:acc_time}
\end{equation}

where \(\xi\) is the magnetic field amplification factor and \(B\) is the upstream magnetic field. For the RS~Oph 2021 (\(\xi B \sim 10^{-7}G\) ), the acceleration timescale for protons up to \(E_p^{\text{cut}} \sim 300GeV\)  is \(t_{\text{acc}} \sim 3.9 d\) .

Meanwhile, the hadronic cooling timescale due to proton-proton interactions, which dominate the hadronic losses in the dense circumstellar environment, is given by,

\begin{equation}
t_{\text{pp}} = (n_p c \sigma_{\text{pp}})^{-1} = 21 \left( \frac{n_p}{6 \times 10^8\ \mathrm{cm}^{-3}} \right)^{-1} \ \mathrm{[d]},
\label{eq:pp_time}
\end{equation}

where \(n_p\) is the target proton density and \(\sigma_{\text{pp}} \approx 30\) mb is the inelastic cross section. In the red giant wind region, \(n_p \sim 6 \times 10^8\) cm\(^{-3}\), so \(t_{\text{pp}} \sim 21~d\).The adiabatic timescale \(t_{\text{ad}} \propto r_{\text{sh}}/v_{\text{sh}}\) is comparable to the expansion timescale during the early phases.

The comparison of these timescales reveals that 
\begin{equation}
t_{\text{acc}} \ll t_{\text{pp}},  \quad t_{\text{acc}} \ll t_{\text{ad}},
\label{eq:timescale_inequality}
\end{equation}

which holds for both the disk and wind environments. This demonstrates that protons can be accelerated to high energies before significant cooling losses occur.

The efficiency of hadronic interactions is quantified by the interaction probability \cite{zhu2021high},

\begin{equation}
f_{\text{pp}} = 1 - \exp\left( - \frac{t_{\text{ad}}}{t_{\text{pp}}} \right).
\label{eq:pp_efficiency}
\end{equation}

During the interaction with the accretion disk, the strong magnetic field significantly reduces the acceleration timescale \(t_{\mathrm{acc}}\), while the high proton density \(n_p\) also shortens the hadronic cooling timescale \(t_{\mathrm{pp}}\). Both timescales are decreased, but \(t_{\mathrm{acc}}\) remains much smaller than \(t_{\mathrm{pp}}\), ensuring efficient acceleration. Moreover, the high density leads to \(f_{\mathrm{pp}} \sim 1\). This implies that nearly all accelerated protons interact with the dense target material, efficiently producing neutrinos and gamma rays. In contrast, in the wind-dominated region, \(n_p\) is lower, resulting in a slower decline of \(f_{\mathrm{pp}}\). 

Energy conversion from the shock kinetic energy into non-thermal particles is assumed to follow a fixed fraction \(\varepsilon_p \sim 0.2\) \cite{MAGIC:2022rmr}. This fraction is consistent with the standard diffusive shock acceleration theory and is also supported by multi-wavelength observations. The proton luminosity \(L_p\) is then converted into gamma-ray and neutrino luminosities via proton-proton interactions.}

We establish an initial geometric model of the evolution of the shock over time, as shown in Fig. \ref{fig:structure}. The key parameters of the system, such as the WD mass, the RG mass loss rate, and the accretion disk mass are adopted based on observations and studies of RS Oph and T CrB.

SymRNe possess a more unique accretion disk structure compared to other novae, enabling partial survival during outbursts and long-term evolution. We hypothesize that the accretion disk of SymRNe is geometrically thin, whose height is much smaller than its radius R. Since the mass of the accretion disk is much smaller than that of WD $M_{\text{disk}} \ll M_{\text{WD}}$, the self-gravity of the disk is negligible. The disk is a flat, axisymmetric structure that surrounds the WD, with its thickness gradually increasing from the inner to the outer regions and featuring a certain opening angle of \( 13.17^\circ \) \cite{Lechuga:2025jei}.

We restrict the present analysis to the free-expansion and adiabatic phases of the shock evolution, and do not model the late radiative stage in detail. The shock is approximated as a thin spherical shell containing accelerated protons. Its radius \( r_{sh} \) evolve with time as,

\begin{equation}
r_{sh} = r_{sh,0} (\frac{t}{1~\mathrm{day}})^{2/3},
\label{eq:shock_radius}
\end{equation}

where $r_{sh,0}$ is $4\times 10^{13}~\mathrm{cm}$\cite{razzaque2010high}. The target particle involved in the hadronic interactions includes two components. For the accretion disk, the proton number density \( n_{\text{disk}} \) is calculated using the adopted thin-disk prescription as in \cite{Lechuga:2025jei},

\begin{equation}
n_{\text{disk}} = 3.1 \times 10^{-8} \beta^{-\frac{7}{10}} \dot{M}_{16}^{\frac{11}{20}} M_1^{\frac{5}{8}} R_{10}^{-\frac{15}{8}} f^{\frac{11}{5}} m_p^{-1},
\label{disk_density}
\end{equation}

where \( m_p \) is the proton mass, \( \dot{M} \) is the accretion rate, and \( R \) is the radial coordinate in the accretion disk. The relevant dimensionless parameters are defined as follows:$\dot{M}_{16} = \frac{\dot{M}}{10^{16} \, M_\odot \, \text{s}^{-1}}$, $M_1 = \frac{M_{\text{WD}}}{M_\odot}$, $R_{10}= \frac{R}{10^{10} \, \text{cm}} $,  $f = \sqrt[4]{1 - \sqrt{\frac{R_{\text{in}}}{R}}}$. The densities of the ejecta \( n_{ej} \) and the RG wind \( n_{RG} \) decrease with increasing shock radius according to
\begin{equation}
n_{\rm ej}=\frac{M_{\rm ej}}{4\pi hr_{\rm sh}^{3}m_{p}}, \qquad
n_{\rm RG}=\frac{\dot{M}_{\rm RG}}{4\pi r_{\rm sh}^{2}v_{\rm RG}m_{p}}.
\label{ej_RG_density}
\end{equation}

Here \( M_{ej} \) is the mass of the ejecta and \( \dot{M}_{RG} \) is the mass loss rate of the RG. The speed of the RG wind is $v_{RG}$. A large number of particles are compressed into a thin shell of thickness $hr_{sh}$ (coefficient h=0.1), resulting in a high density in the shell. Since the opening angle of the accretion disk is limited, only a fraction of the shock surface interacts directly with the disk, while the remaining part propagates through the RG wind.

\subsection{Proton spectrum and neutrino production }

The high-energy gamma-ray and neutrino emission is assumed to originate from hadronic interactions between shock-accelerated protons and the surrounding target material. The parent proton population is described by a time-dependent non-thermal spectrum of the form \cite{DeSarkar:2023nhp},

\begin{equation}
\frac{\partial N_p}{\partial E_p} \propto E_p^{-\alpha_p} \exp\left(-\frac{E_p}{E_p^{\text{cut}}}\right),
\end{equation}

where \( \alpha_p\) is the spectral index, and \( E_p^{\text{cut}} \) is the cutoff energy of the accelerated protons. This parametrization is intended to account effectively for particle acceleration on the shock, together with the effects of cooling and escape.
The normalization of the proton spectrum is determined through the proton luminosity,

\begin{equation}
L_p = \int E_p \frac{\partial N_p}{\partial E_p} dE_p.
\label{eq:proton_luminosity}
\end{equation}

Once the proton distribution and the target density are specified, the hadronic gamma-ray spectrum can be calculated using the standard semi-analytical treatment of proton-proton interactions as \cite{Kelner:2006tc}. Where neutral and charged pions are produced simultaneously. Neutral pions decay into gamma rays, whereas charged pions decay through muons and produce neutrinos,
\begin{equation}
\pi^{0}\rightarrow \gamma+\gamma,~~~ \\
\pi^{\pm}\rightarrow \mu^{\pm}+\nu_{\mu}(\bar{\nu}_{\mu})
\rightarrow e^{\pm}+\nu_{e}(\bar{\nu}_{e})+\nu_{\mu}+\bar{\nu}_{\mu}.
\end{equation}

When the shock sweeps through the accretion disk, since the opening angle of the accretion disk \( \theta_{\text{disk}} \) is approximately \( 13.17^\circ \) \cite{Lechuga:2025jei}. We calculated the solid angle for interaction between the shock with the disk, which is approximately $ 2\pi\sin\left(\frac{\theta_{\text{disk}}}{2}\right) \sim 0.75 $ (taking about 6\% of the whole region), and in the other regions the shock interacts only with the RG wind.

For hadronic process, the energy ratio of produced neutrino and gamma rays is $\sim 1/2$. The total neutrino luminosity is approximately twice the gamma-ray luminosity \cite{Halzen:2019qkf}, while the observable muon-neutrino component at Earth is reduced by flavor oscillations to about one third of the total neutrino flux, $L_{\nu }\approx (2/3)L_{\gamma}$. 

After the shock has crossed the dense disk region, it continues to propagate mainly through the RG wind. {\color{black}
The evolution of shock can be characterized by the five critical stages: forming of shock, beginning of the interaction with shock and disk, beginning of shock deceleration, transition adiabatic phase into radiative phase, and fully radiative phase \cite{Zheng:2024qwt}. }{\color{black}When the swept-up mass is equivalent to the ejecta mass,} the shock decelerates significantly and transitions from the adiabatic phase to the radiative phase. {\color{black}This transition is governed by the competition between the adiabatic timescale (\(t_{\text{ad}}\)) and the radiative cooling timescale (\(t_{\text{cool}}\)), quantified by the radiation efficiency parameter \(\epsilon_{\text{rad}}\).}
In the adiabatic phase ($\epsilon_{\mathrm{rad}} = 0$), the shock decelerates slowly, free-free emission dominates, and the gamma-ray flux follows $F_\gamma \propto t^{-4/3}$. As the shock decelerates further, the temperature drops and the swept-up mass increases, causing the radiative cooling efficiency to rise sharply. {\color{black}When the radiative cooling timescale (\(t_{\text{cool}}\)) becomes comparable to the adiabatic timescale (\(t_{\text{ad}}\))}, the shock enters the radiative phase $\epsilon_{\mathrm{rad}} > 0$), where line radiation dominates and the kinetic energy of newly swept-up material is radiated almost instantly. The shock then decelerates more rapidly, and the flux decay steepens to $F_\gamma \propto t^{-2/3}$. The fully radiative stage corresponds to $\epsilon_{\mathrm{rad}} = 1$. In this way, the temporal evolution of the hadronic emission provides a direct probe of the shock dynamics and of the density structure of the circumstellar medium.

\section{The case of RS Oph } \label{Application to RS Oph}
RS Oph is one of the best-studied SymRNe in our Galaxy and is therefore a suitable benchmark source for the present model. Its recurrence timescale is approximately (15\text{--}20) yr due to the high accretion rate. Its most recent eruption in 2021 August was extensively observed over a broad energy range. Thanks to the availability of multi-wavelength observations, RS Oph provides an appropriate system for testing the time-dependent emission scenario developed in this work.

\vspace{1cm}
\begin{table*}[htbp]
\begin{tabular}{ccccc}
\hline
\textbf{Parameter} & \textbf{Symbol} & \textbf{Value} \\ \hline
  WD's mass & $M_{\text{WD}}$ & $1.37M_\odot$ \\
  Accretion rate & $\dot{M}$ & 2\texttimes$10^{-7}M_{\odot}/yr$ \\
  Inner radius of accretion disk & $R_{\text{in}}$ &$16.6R_{\odot}$ \\
  Range coefficient & $\beta$ & 0.3 \\
  Mass loss rate of the RG & $\dot{M}_{\text{RG}}$ & 5\texttimes$10^{-7}M_{\odot}/yr$\\
  Speed of the RG wind & $v_{\text{RG}}$ & $10 km/s$\\
  Mass of the ejecta & $M_{\text{ej}}$ &$10^{-6}M_\odot$ \\
  Cutoff energy of accelerated protons & $E_p^{\text{cut}}$ & 250-800 GeV \\
  Spectral index & $\alpha_p$ & 2.4 \\
\bottomrule
\end{tabular}\caption{Adopted parameters of the RS Oph 2021 \cite{MAGIC:2022rmr, shara2018masses, Kelner:2006tc} }
    \label{Table 3}
\end{table*}

\subsection{Adopted parameters}
The physical parameters adopted for RS Oph are summarized in Table~\ref{Table 3}. These include the WD mass, the accretion rate, the properties of the red giant wind, the ejecta mass, and the parameters describing the proton distributioin. In particular, the WD mass is taken to be 1.37 $M_{\odot}$, the accretion rate is $2\times10^{-7} M_{\odot} \mathrm{yr^{-1}}$\cite{Lechuga:2025jei}, the red giant mass-loss rate is $5\times10^{-7} M_{\odot}~\mathrm{yr^{-1}}$, and the wind velocity is $10~\mathrm{km~s^{-1}}$. The ejecta mass is assumed to be $10^{-6} M_{\odot}$\cite{MAGIC:2022rmr}.

The parameters of the proton spectrum are constrained through fits to the high-energy data of the 2021 eruption. With GAMERA \cite{2015ICRC...34..917H} we fit the gamma-ray observations and find the spectral index is kept constant over the first four days at 2.4, while the cutoff energy is evolved with time, taking representative values of 250, 500, 600, and 800 GeV for the first four days \cite{DeSarkar:2023nhp}, respectively and the luminosity of proton is $3.4\times 10^{38}~\mathrm{erg s^{-1}}$. These values are then used to calculate the corresponding hadronic gamma-ray and neutrino emission within the framework described previously.

\subsection{Optical calibration}
\begin{figure}
    \centering
    \includegraphics[width=0.45\textwidth]{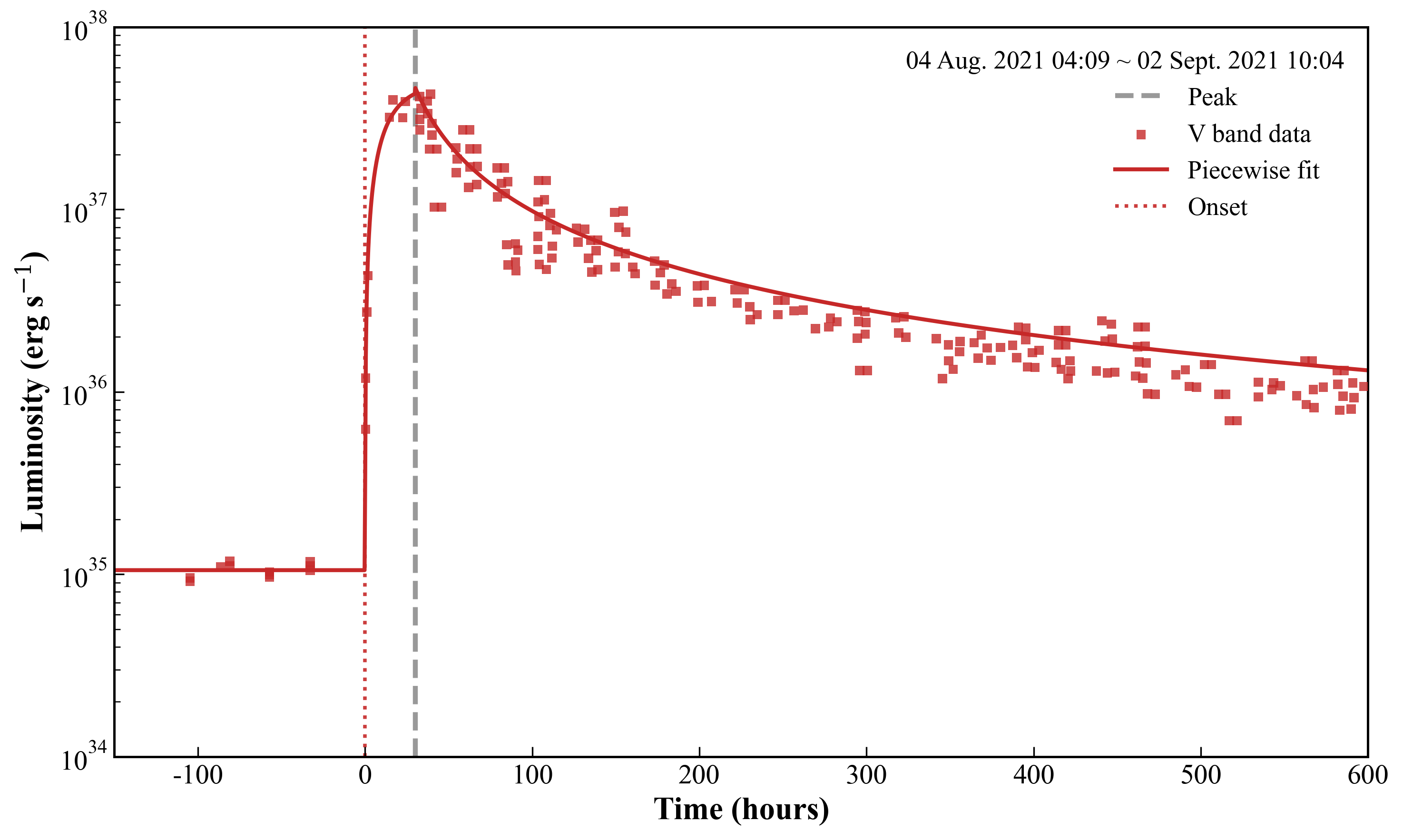}
    \caption{Optical data (red points) and fitting curve (solid line) of RS Oph from 4th August to 2nd September 2021. The peak of optical luminosity is about 27.6 h after the starting point of the eruption. The onset and peak time are presented with red dotted and grey dashed lines.}
    \label{fig:RS Oph 2021 optical fit}
\end{figure}

We take the V-band optical data of 2021 eruption from AAVSO, as shown in Fig. \ref{fig:RS Oph 2021 optical fit}. We find the the rise time $t_{rise}$ of the optical data is about 27.6 hours. The accreted layer on the WD surface is taken to be geometrically thin, the thickness $\varDelta R$ is estimated about 274 km as following,
\begin{equation}
\varDelta R = \sqrt {c~{t}_{rise}~\lambda}, 
    \label{equ:diff_time}
\end{equation}

where $\lambda$ is the mean free path $\lambda=\frac{1}{\kappa \rho}$ and $c$ is the speed of light. We also adopt a conservative lower-limit opacity of $\kappa =0.1 {cm}^{2}/g$ for the envelope, consistent with typical estimates of free-free opacity in stellar matter \cite{prialnik2009introduction}.

\begin{figure}
    \centering
    \includegraphics[width=0.45\textwidth]{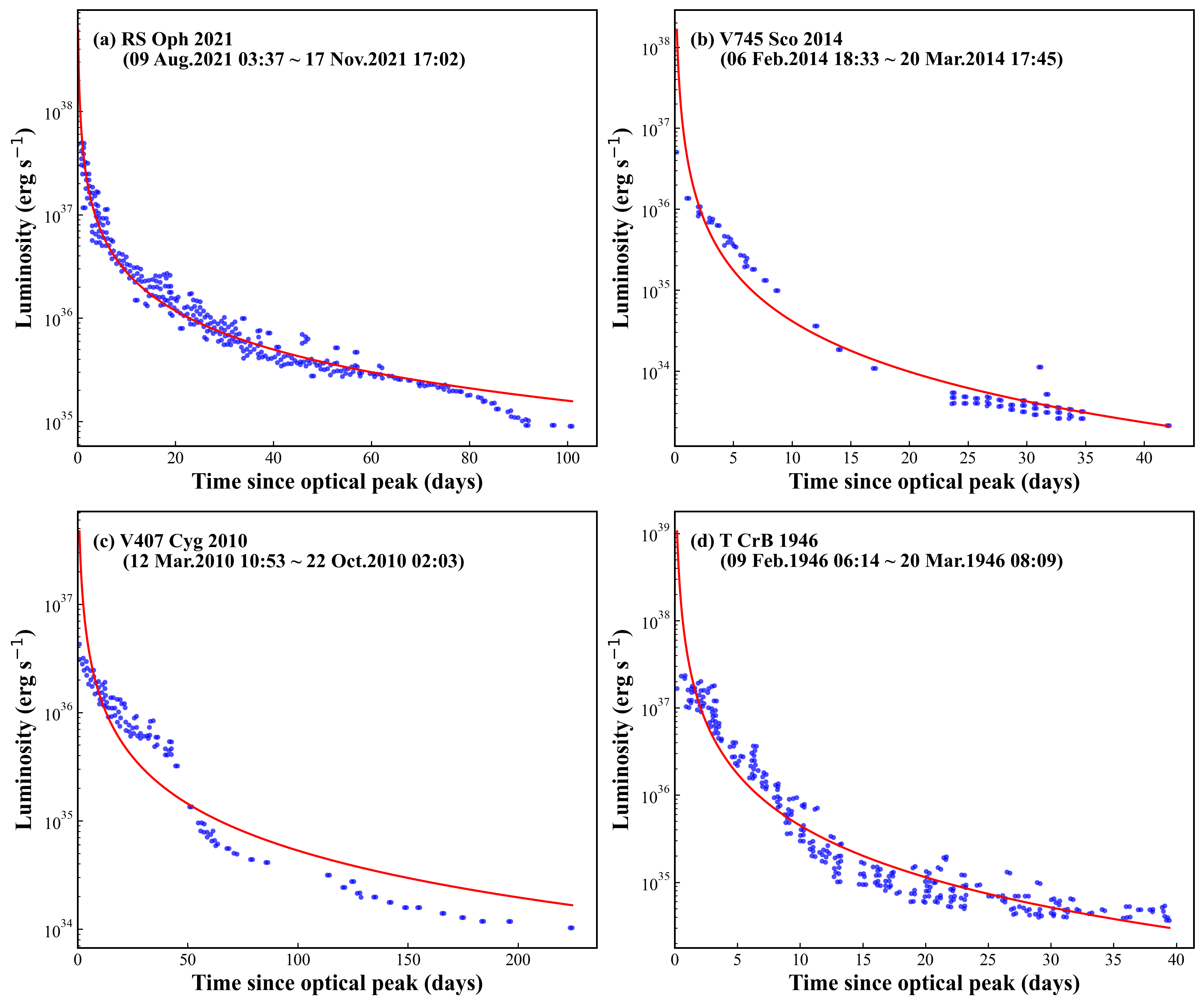}
    \caption{V-band data (blue dots) and fitting curve (red solid line) of RS Oph 2021, T CrB 1946, V407 Cyg 2010 and V745 Sco 2014. Data is taken from AAVSO\footnote{https://www.aavso.org/LCGv2/}. }
    \label{fig:four novae lightcurves combined}
\end{figure} 

After the optical peak, the light curve gradually decays. For comparison, we show the optical light curves of RS Oph, T CrB, V407 Cyg, and V745 Sco in Fig.~\ref{fig:four novae lightcurves combined}. The light curves are fitted using a power-law form, ($L_{opt}\propto t^{-\alpha}$), where \(L_{opt}\) denotes the V-band luminosity and \(\alpha\) characterizes the post-peak decay rate. This comparison provides a phenomenological way to assess differences in the circumstellar environment and ejecta evolution among these systems.
We list the main factors influencing SymRNe optical signals and their fitting values in Table \ref{tab:optical signals for SymRNe}. There exists a close connection between the mass-loss rate of the RG, orbital period, and the accretion rate to the WD. A large amount of material from the RG is gravitationally captured by the WD. In general, a higher mass-loss rate leads to a larger amount of accreted mass. However, during the accretion process, some material is inevitably lost; only a fraction of the transferred material actually reaches the accretion disk and the surface of WD. Differences in the mass lost lead to variations in the amount of accumulated mass, resulting in different outburst recurrence periods.
Compared to the other three SymRNe, V407 Cyg has a relatively low WD mass. Additionally, its unstable Mira RG introduces large uncertainties in its accretion rate \cite{hachisu2018light}. Therefore, based on its outburst recurrence period of $\sim70$ yr and ejected mass ($\sim10^{-6}\,M_{\odot}$ yr$^{-1}$), we estimate an average accretion rate.

\begin{table*}[htbp]
\begin{tabular}{cccccc}
\hline
SymRNe  & \begin{tabular}[c]{@{}c@{}}Attenuation \\index $\alpha$   \end{tabular}    & \begin{tabular}[c]{@{}c@{}}Decay time\\  (day)    \end{tabular}    & \begin{tabular}[c]{@{}c@{}}RG Mass loss\\   rate ($10^{-7}M_\odot$/yr) \end{tabular}   &  \begin{tabular}[c]{@{}c@{}}Average accretion\\  rate ( $10^{-7}M_\odot$/yr) \end{tabular}  & \begin{tabular}[c]{@{}c@{}}The peak luminosity\\($10^{38}erg/s$)  \end{tabular} \\ \hline
RS Oph 2021  & \begin{tabular}[c]{@{}c@{}}1.244\end{tabular} & 102    & $5^{\cite{DeSarkar:2023nhp}}$ & $0.7-2.0^{\cite{Lechuga:2025jei,shara2018masses}}$ & 21.4 \\ T CrB 1946 & 1.969 & 38 & $5^{\cite{Wang:2025ibu}}$  & $0.2-0.4^{\cite{Wang:2025ibu,shara2018masses}}$ &6.9  \\ V407 Cyg 2010  & 1.437   & 220     & $3^{\cite{razzaque2010high}}$  & 0.14 &0.3 \\ V745 Sco 2014   &  2.079   &  42       & $1^{\cite{Banerjee:2014eua}}$   &  $0.11^{\cite{shara2018masses}}$ & 2.6   \\ \hline
\end{tabular} 
\caption{Parameters related to the optical signals for SymRNe. }
    \label{tab:optical signals for SymRNe}
\end{table*}

Their optical decay timescales differ substantially. In particular, RS Oph and V407 Cyg show relatively slow declines, with decay indices smaller than about 1.5, and require several hundred days to return to quiescence. By contrast, V745 Sco and T CrB are fade much more rapidly, with decay indices close to 2.0 and characteristic decay times of only a few tens of days. 
The decay time is strongly affected by the density and spatial distribution of the circumstellar medium (CSM), but the relationship is not simply monotonic. A denser CSM enhances shock interaction and can lead to rapid early decline, it also provides a larger reservoir of material that can be ionized and later recombine, producing a prolonged optical tail. 

For RS Oph, the relatively slow optical decay is consistent with a dense CSM and strong interaction between the ejecta and the surrounding material. Its short recurrence period implies a relatively high accretion rate and the formation of a comparatively massive circumstellar envelope around the WD \cite{RS_oph}. The resulting shock interaction can efficiently deposit energy into the ambient medium, thereby contributing to a prolonged decline in the optical band. In contrast, V745 Sco and T CrB do not possess such dense RG winds. They have higher WD mass and smaller initial envelope mass. Therefore, the critical mass triggering the TNR is relatively small. When the nova erupts, it directly enters the rapid decay phase \cite{light_curve}. V407 Cyg shows a much longer evolution, consistent with its extended Mira-type circumstellar environment \cite{V407}.

\subsection{Time-dependent MeV neutrinos}
\begin{figure}
    \centering
    \includegraphics[width=0.45\textwidth]{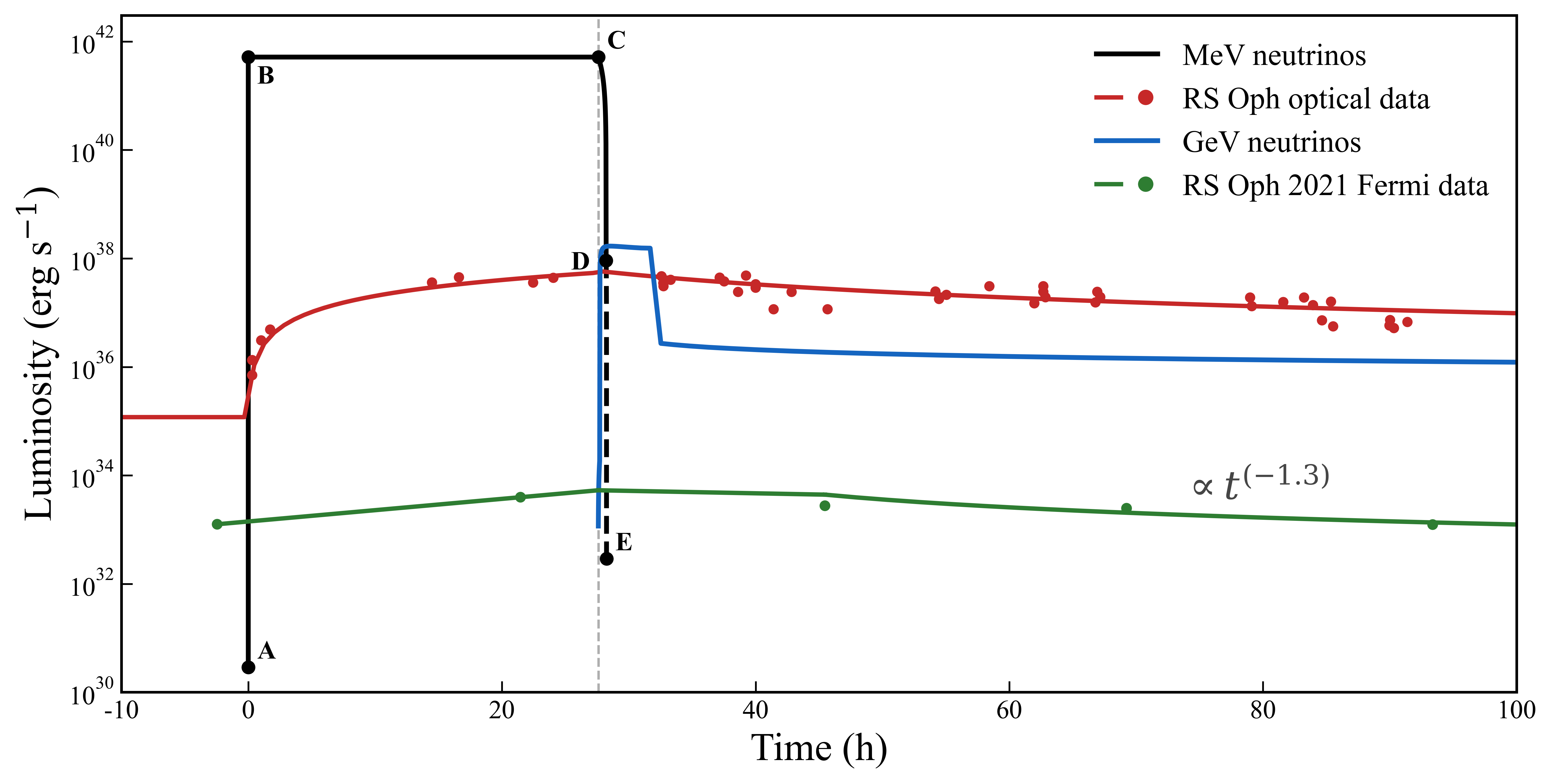}
    \caption{RS Oph 2021 optical data and fitting curve (red), RS Oph 2021 high-energy band data and fitting curve observed by \textit{Fermi}-LAT (green), estimated MeV neutrino light curve generated by TNR (black), estimated GeV neutrino light curve generated by hadron processes (blue).}
    \label{fig:RS Oph comprehension neutrinos and data.png}
\end{figure}

The time evolution of the MeV neutrino is constructed on the basis of the thermonuclear evolution discussed in Sect. \ref{Physical scenario and model setup}. In particular, we refer to the work of Wang et al. \cite{Wang:2025ibu}, who simulated the evolution of an ONeMg WD during a nova eruption. We adopt a peak low-energy neutrino luminosity of $5.16\times 10^{41}~\mathrm{erg/s}$. The MeV neutrino light curve is parameterized in terms of five characteristic stages, labeled Point A to E in Fig. \ref{fig:RS Oph comprehension neutrinos and data.png}.

Point A corresponds to the onset of the TNR. The low-energy neutrino and gamma rays are generated.From Point A to B, the temperature rises rapidly and the nuclear reaction rate increases sharply. Neutrinos freely escape from the accreted lay, leading to an immediate luminosity increase. 

Point B is defined as the moment when photons begin to emerge from the surface of the accreted layer as shown in Fig. \ref{fig:RS Oph comprehension neutrinos and data.png}. At this stage, the nuclear burning efficiency reaches its maximum and the MeV neutrino luminosity attains its peak value. We define this point as the starting time ($T_0$). Between Point B and C, the nuclear reactions remain at high efficiency and the MeV neutrino luminosity remains close to its maximum, while the optical emission is still rising toward its peak because of the delayed escape of photons through the envelope.

Point C marks the epoch at which the optical luminosity reaches its maximum. According to our optical fitted results for RS Oph 2021 in Fig. \ref{fig:RS Oph 2021 optical fit} , this occurs about 27.6 hours after the $T_0$ of the eruption. 
From Point C to D, the temperature reaches $7\times10^{7}K$, the accumulated thermal pressure exceeds gravitational binding, degeneracy becomes less significant, and the shell begins to expand outward \cite{starrfield2016thermonuclear}, forming the shock, shortly starting to interact with disk and stellar wind. We define Point C as $T_1$.
As the envelope expands, the density and temperature in the nuclear burning region decreases, reducing the reaction rate and causing the MeV neutrino luminosity to decline super rapidly. In contrast, the photon luminosity remains approximately constant for some time because the ejecta are still optically bright.

Point D is associated with the stage at which the expanding shell reaches a radius of about $10^{12} cm$, and most of the released energy has been converted into shock kinetic energy, and the nuclear reaction rate rapidly decreases. We expect this occurs about 0.62 hours after the onset of shell expansion, assuming an initial shock velocity of about 4500~$\mathrm{km~s^{-1}}$. 
From Point D to E, the WD photosphere begins to contract, the nuclear burning weakens further, and the low-energy neutrino luminosity remains at a much reduced level until the end of the burning phase. In this way, the MeV neutrino light curve traces the thermonuclear evolution more directly than the optical light curve, especially during the earliest stages of the outburst.

Point E indicates the end of nuclear reactions. From Point E to A, the nuclear reaction process is over, and the luminosity rapidly declines.
Based on the full cycle described above, we plot the flux of low-energy neutrinos in Fig. \ref{fig:four novae lightcurves combined} during a nova outburst. 

\subsection{Time-dependent GeV emission}
The GeV neutrino emission is calculated by combining the shock evolution described in Sect. \ref{Physical scenario and model setup} with the time-dependent target densities in the ejecta, the accretion disk, and the red giant wind. For each time $t$, the corresponding values of $E_p^{\rm cut}$ (obtained with GAMERA), $n_{\text{RG}}$, $n_{\text{ej}}$, $n_{\text{disk}}$ are evaluated and the proton spectrum is used to derive the hadronic gamma-ray emission and the associated neutrino output. The resulting GeV neutrino luminosity is shown in Fig. \ref{fig:RS Oph comprehension neutrinos and data.png}.

The model predicts that the GeV neutrino emission rises rapidly when the shock begins to interact with the dense accretion disk. Because the disk density is much higher than that of the RG wind, the hadronic interaction rate increases strongly at this stage, producing an early peak in the GeV neutrino luminosity. When the shock has crossed the disk region, at approximately 4.56 hours with the disk, the effective target density decreases abruptly, and the neutrino luminosity drops. Later, the shock interacts mainly with the RG wind, and the GeV emission decays more gradually. 
In the symbiotic binary system, the WD resides in close proximity to the RG. Dense wind material, originating from the RG, permeates the region surrounding the WD, creating an extensive CSM. According to simulations \cite{Booth:2016bga}, the spatial distribution of this wind material extends over a significant scale, ranging from approximately $\pm (1.1-2.2)\times 10^3 R_\odot$. The density profile of the wind is governed by the radial distance from the WD. As described by Eq. \ref{ej_RG_density}, the wind density is highest in the vicinity of the WD (the center of the shock interaction) and decreases as the shock radius $r_{sh}$ increases. When a nova outbursts, high-velocity ejecta are launched from the surface of the WD. These ejecta expand into this pre-existing dense medium, and are heavily influenced by this dense wind, until the shock eventually expands beyond the confinement of the RG wind.

A notable feature of the multi-energy signal is the temporal offset between the MeV and GeV components. The MeV neutrino luminosity reaches its peak before the optical maximum and starts to decline significantly after Point C, whereas the GeV component rises shortly thereafter as the shock develops and interacts with the circumstellar material. The overall neutrino light curve therefore exhibits a two-stage structure, with an early thermonuclear component followed by a delayed hadronic component. This behavior provides a direct link between the underlying nuclear burning, the shock evolution, and the density structure of the immediate nova environment.

During the evolution of the shock, the propagation of high-energy gamma rays through the medium is subject to absorption, resulting in an observed flux much lower than the intrinsic radiation flux. As $\gamma$ rays typically interact with protons or the radiation field and decay into electron-positron pairs. The optical depth \(\tau\) can be calculated as follows \cite{Petruk:2026eeg},
\begin{equation}
\tau = \int \mu \, dL,
\end{equation}
where \(\mu\) is the absorption coefficient, depending on the density of the ambient medium, and \(L\) is the propagation length. For \(\gamma p\) absorption, the absorption coefficient ($\mathrm{{cm}^{-1}}$) can be expressed as \cite{berezinskii1984astrophysics},
\begin{equation}
\mu_{\gamma p}(E_\gamma) \approx 3.6 \times 10^{-27} n_p \left[ \ln \left( \frac{E_\gamma}{m_e c^2} \right) - 1.9 \right] \quad
\end{equation}
where \(n_p\) is the proton number density, \(m_e\) is the electron mass, and \(E_\gamma\) is the gamma-ray energy. Since the proton number density in the RG wind is much lower than that in the accretion disk, we neglect the wind contribution and compute only the optical depth \(\tau_{\gamma p}\) within the accretion disk.

Based on the constructed model, the average density of the accretion disk is \(\sim 5 \times 10^{12} \, \text{cm}^{-3}\), and the propagation length is taken as the disk extent \(\sim 174~R_\odot\). For a 1 GeV gamma ray, the optical depth \(\tau_{\gamma p}\) is approximately 4.1. In addition, absorption due to the nova's own photon field (\(\gamma\gamma\) interaction) is also important. For a 1 TeV gamma ray, the absorption coefficient is:
\begin{equation}
\mu_{\gamma\gamma}(E_\gamma) \approx 7 \times 10^{-26} \omega_\gamma \quad
\end{equation}
where \(\omega_\gamma = 3L/(4\pi R^2 c)\) is the average radiation energy density, and \(L\) is the bolometric luminosity. For typical nova luminosities (\(\sim 10^{38} \, \text{erg/s}\)), the optical depth \(\tau_{\gamma\gamma}\) is about 2.53. Clearly, such large optical depths prevent most high-energy gamma rays from escaping. The contribution of the accretion disk is not significant at the highest gamma-ray energies. The gamma-ray luminosity fit with $t^{-1.3}$, consistent with the theoretical study.

\section{Detectability and implications} \label{Detectability and implications}
\subsection{MeV neutrino spectra}
The MeV neutrino emission is associated with nuclear and thermal processes during the TNR phase \cite{baxter2022snewpy}. The nuclear reactions during the TNR phase of a nova are dominated by the CNO cycle, and the electron neutrino spectrum from \(\beta^+\) decays follows a Gamma distribution with a mean neutrino energy \cite{Wang:2025ibu}. As seen in Fig.~\ref{fig:RS Oph comprehension neutrinos and data.png}, the MeV neutrino luminosity remains the peak from Point B to C. Although the luminosity rises rapidly from A to B, this phase is too short to be observationally accessible. After point C, the luminosity drops sharply, which is below the detectable levels. Consequently, we adopt the Point B to C as the optimal observational time window and use the corresponding luminosity as the normalization factor to estimate the MeV neutrino spectrum.
The specific spectral formula is,
\begin{equation}
\Phi(E_\nu) = \frac{L_\nu}{4\pi D^2 \langle E_\nu \rangle} \left( \frac{\alpha_M + 1}{\langle E_\nu \rangle} \right)^{\alpha_M} \exp\left(-\frac{(\alpha_M + 1)E_\nu}{\langle E_\nu \rangle}\right),
\end{equation}

where \(L_\nu\) is the peak MeV neutrino luminosity, about $5.16\times 10^{41}erg/s$. Considering the consistency exhibited by the four SymRNes in various physical parameters, we assume that their MeV neutrino peak luminosity is the same. According to \cite{BOREXINO:2020aww}, although the temperature and density of a nova differ from those in the solar core, the mean energy of CNO-cycle \(\beta\)-decay neutrinos remains in the MeV range. Based on the CNO fusion sequence and the solar neutrino energy spectrum, the average value lies between 1.1 and 1.7 MeV, we adopt \(E_\nu\) = 1.3\ \text{MeV}\ as the mean energy of CNO cycle neutrinos. For the spectral index, we take \(\alpha_M = 2.2\), which is suitable for neutrino spectra generated by thermonuclear burning \cite{Keil:2002in}.

\begin{figure}
    \centering
\includegraphics[width=0.5\textwidth]{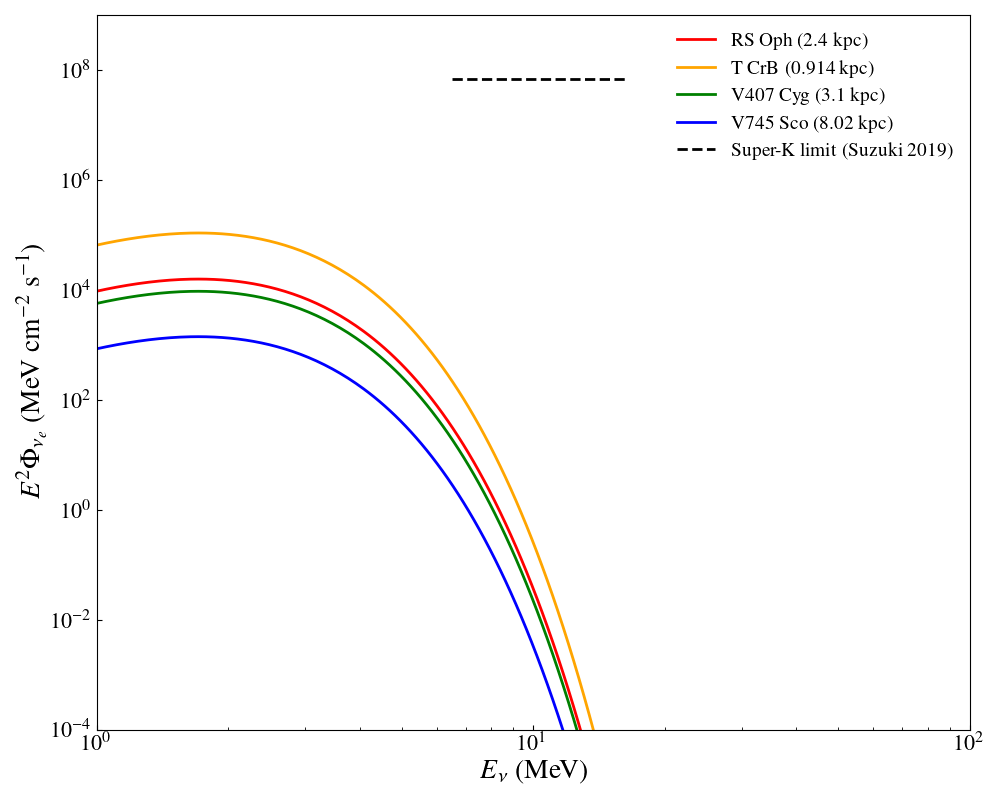}
    \caption{The comparison of MeV neutrino spectrum reaching the Earth from RS Oph, T CrB, V407 Cyg, and V745 Sco (red, orange, green and blue solid line). The detection upper limit of Super-Kamiokande on electron neutrino flux (black dotted line).}
    \label{fig:nutrinos flux MeV}
\end{figure}

We show the MeV neutrino spectra of four SymRNe and upper limit of Super-Kamiokande \cite{Suzuki:2019jby} in Fig. \ref{fig:nutrinos flux MeV}, which is inversely proportional to $D^2$. Unfortunately, the MeV neutrino flux remains well below the detection threshold of current detectors. 
This is mainly due to its relatively low flux and the limited sensitivity of existing detectors in the relevant energy range. Even for nearby sources such as RS Oph, the expected event rate is too small to yield a statistically significant signal. Therefore, the MeV component should primarily be regarded as a theoretical diagnostic of the thermonuclear processes rather than an observational target with current technology.

\subsection{GeV detectability}
By incorporating the accretion disk into our model, we find that it can significantly modify both the spectral shape and the temporal evolution of the neutrino emission during the early phase.

\begin{figure}[htbp]
    \centering
    
    \begin{subfigure}{0.9\linewidth}
        \centering
        \includegraphics[width=\linewidth]{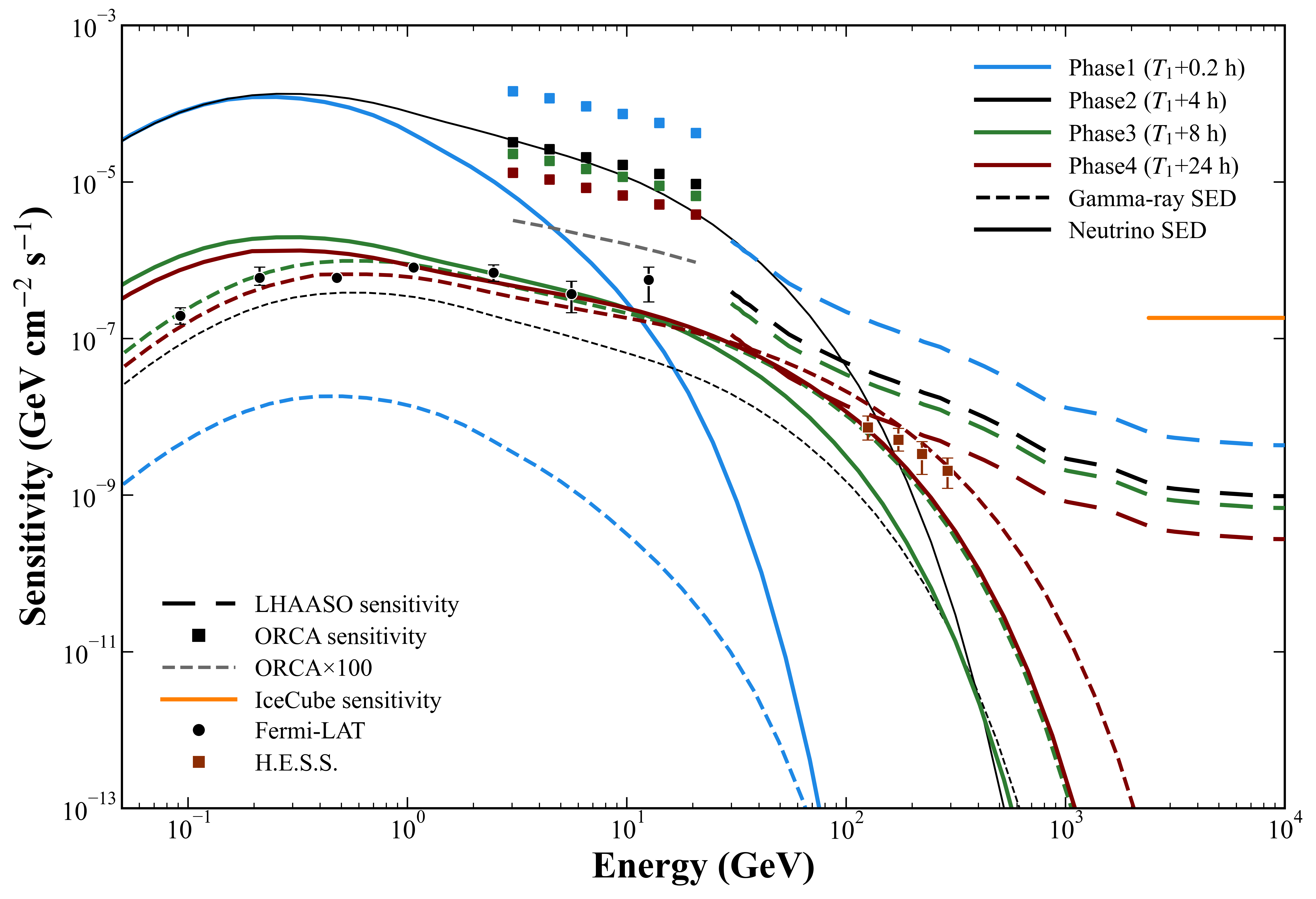}
    \end{subfigure}
       
    \begin{subfigure}{0.9\linewidth}
        \centering
        \includegraphics[width=\linewidth]{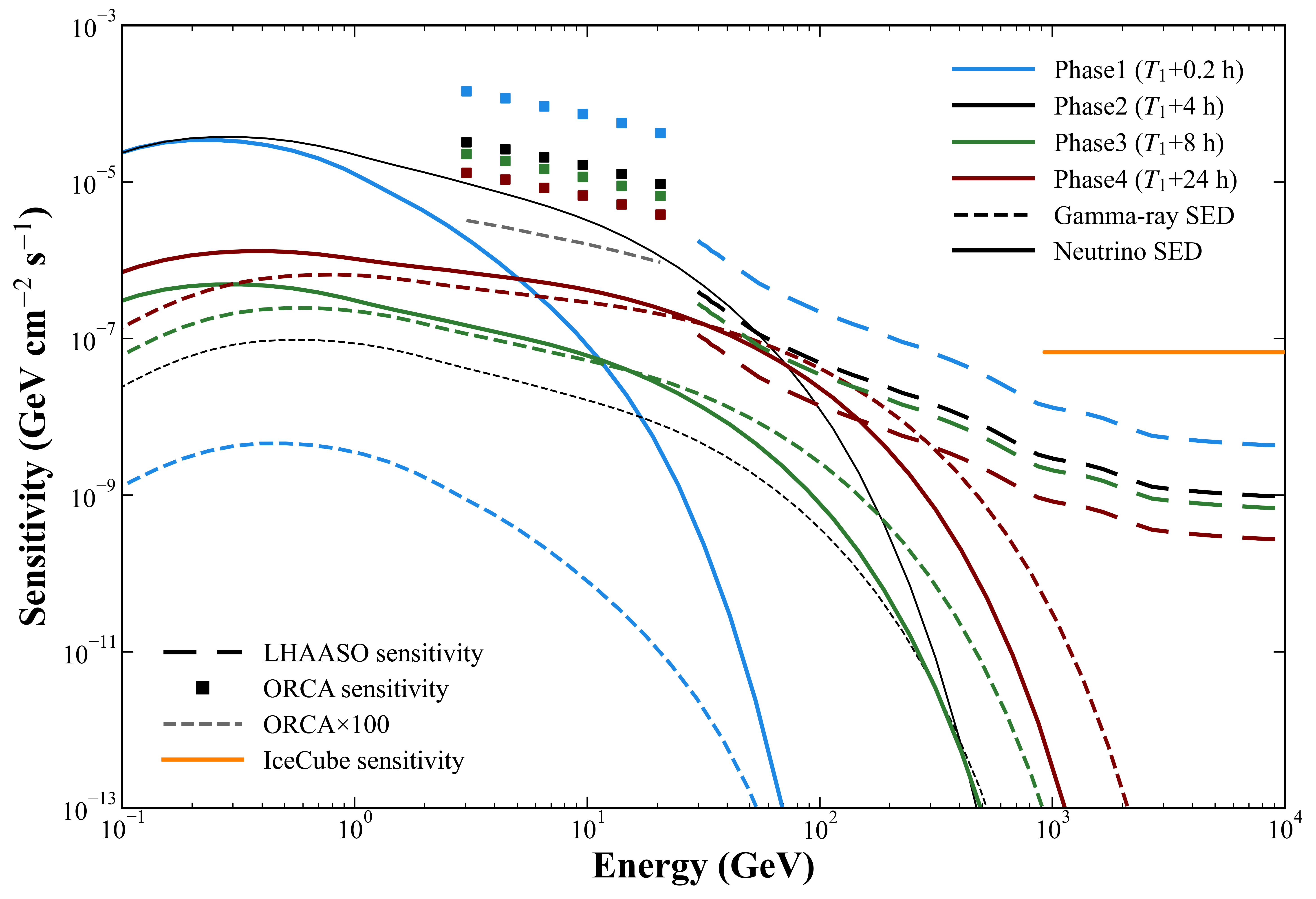}
    \end{subfigure}
    
    \caption{The GeV neutrino (solid line) and gamma rays (dashed line) spectra at four phases during the early stages of RS Oph (top) and T CrB (bottom). The detection limits of ORCA (squares), ORCA$\times 100$ (grey dashed){\color{black},IceCube(orange) }and LHAASO (long-dashed) are marked. The neutrino spectra, ORCA sensitivity, LHAASO sensitivity for Phase 1-4 are in blue, black, green and red respectively. The observations by \textit{Fermi}-LAT \cite{MAGIC:2022rmr} and H.E.S.S. \cite{HESS:2022qap} are labeled as blue circle and red squares.}
    \label{fig:combined}
\end{figure}

In order to describe the spectral evolution at different times and study the role of accretion disks, we introduced four specific phases, particularly focusing on the stage of shock disk collision. At the moment of shock formation ($T_1$), the interaction with the disk lasts approximately 4.56 h. For $t \lesssim T_1+0.2h$, the shock has not yet encountered the disk, and the neutrino emission is dominated by the RG wind. As illustrated in Fig.~\ref{fig:combined}, at Phase~1 ($T_1 + 0.2h$) the shock begins to collide with the accretion disk with the highest target particle density. It leads to dramatic increase in neutrino emission; however gamma rays are absorbed severely. As the shock propagates outward, both the gamma-ray and neutrino fluxes increase significantly, and the maximum energy of protons extends to higher values. In Phase~2 ($T_1 + 4h$), the shock is about to sweep through the disk; the neutrino flux remains high, but the gamma-ray flux is still significantly lower than the neutrino flux due to gamma-ray absorption by the accretion disk. In Phase~3 ($T_1 + 8h$), the shock has swept through the disk, the role of the disk vanishes, and the target proton density drops. The emission is again dominated by the RG wind, the neutrino flux declines and approaches the gamma-ray flux. Despite significant fluctuations in neutrino flux in the early stages, after Phase~4 ($T_1 + 24h$), the spectrum gradually stabilizes. 

The \textit{Fermi}-LAT \cite{MAGIC:2022rmr} and H.E.S.S. \cite{HESS:2022qap} observations of high-energy gamma rays from the 2021 outburst of RS Oph is shown, taken on the first day after the outburst (corresponding to one day after $T_1$ in our notation). During the stable spectral phase from Phase~3--4, the predicted gamma-ray flux matches the observations well, validating our model. \textcolor{black}{With increasing outburst duration, the sub-TeV gamma-ray flux rises accordingly. Temporal variability studies of gamma-ray emission can provide constraints on key accretion disk parameters, including its radius and vertical thickness. GeV neutrinos is the dominant emission channel at Phase 1 and 2. For the future T CrB outburst, LHAASO is capable to detect its Phase-4 TeV emission.}

\textcolor{black}{To assess the detectability of the predicted GeV neutrinos, we compute the expected number of signal and background events in the KM3NeT/ORCA detector. We adopt the effective area \(A_{\mathrm{eff}}(E)\) from \cite{deWasseige:2020dnq} and the atmospheric neutrino background flux \(\Phi_{\mathrm{atm}}(E)\) from the Honda model\cite{PhysRevD.92.023004}. The expected signal events in the \(i\)-th energy bin are obtained by direct integration:}
{\color{black}
 \begin{equation}
S_i = T \int_{E_i}^{E_{i+1}} \Phi_\nu(E) \, A_{\mathrm{eff}}(E) \, dE,
\label{eq:signal}
 \end{equation}

 where \(T\) is the exposure time and \(\Phi_\nu(E)\) is the predicted neutrino flux at Earth from our model (shown as solid curves in Fig.\ref{fig:combined}). Similarly, the background events are

 \begin{equation}
B_i = T \int_{E_i}^{E_{i+1}} \Phi_{\mathrm{atm}}(E) \, A_{\mathrm{eff}}(E) \, dE.
\label{eq:background}
 \end{equation}

The total significance is estimated as \(\sigma = \sum_i S_i / \sqrt{B_i}\), and we require \(3\sigma\) for a detection. The predicted number of events for RS~Oph and T~CrB during the four phases are summarized in Table~\ref{tab:the expected numbers of events}. The sensitivity curves shown in Fig.~\ref{fig:combined} represent the flux required for a detection of \(3\sigma\). In Phase~2, the predicted flux is below the ORCA sensitivity curve, and the expected signal events are well below unity. Therefore, a detection with ORCA is not possible during any phase. However, if the effective volume was increased by a factor of 100 (ORCA\(\times\)100), the signal events would scale proportionally, making a \(3\sigma\) detection feasible for Phase~2.

In Fig.~\ref{fig:combined}, the predicted neutrino spectra (solid curves) are shown together with the \(3\sigma\) sensitivity curves of ORCA (black dotted), ORCA\(\times\)100 (grey dashed), and LHAASO (long-dashed). The enhancement provided by the accretion disk is visible in Phase~2 (black solid curve), but it is not pronounced. The sensitivity of ORCA\(\times\) 100 is systematically lower, and the predicted Phase~2 spectrum crosses it, indicating that a detector with a 100 time larger effective area would be able to claim the observation.
}

We show the GeV gamma-ray spectra of RS~Oph and T~CrB in Fig.~\ref{fig:combined}. For late-phase, such as Phase 3 and 4, protons are accelerated to higher energies, with GeV - TeV gamma-ray emission, LHAASO should be able to detect both sources~\cite{bai2019large}.{\color{black}For comparison, we also show the IceCube sensitivity curve in Fig.~6 \cite{IceCube:2025egb}. Since IceCube is primarily optimized for the TeV-PeV energy range, its sensitivity lies well above our predicted neutrino flux, making the detection of the early-time GeV signal extremely challenging. This is consistent with the non-detection of correlated neutrinos from novae by IceCube~\cite{IceCube:2025egb}.}

\subsection{Application to T CrB }
Although T CrB and RS Oph have some features in common, T CrB has lower peak luminosity, and exhibited a noticeably faster decline than RS Oph as shown in Fig. \ref{fig:four novae lightcurves combined}. It had two eruptions in 1866 and 1946. The plausible range of rise time is between 2.8-12 hours \cite{T_CRB_rise_time}, so here we adopt 6 hours as our input.  

The shape of MeV neutrino light curve of T CrB is similar to that of RS Oph, since both are mainly governed by the nuclear and thermal reactions, but with shorter emission time. In contrast, the GeV neutrino light curve shows a mild but non-negligible difference.  
The expected GeV neutrino spectrum is presented in Fig.~\ref{fig:combined}. Because its lower accretion rate and less dense RG wind, T CrB yields a lower neutrino flux than RS Oph. Even for the most promising Phase, its flux is still one order of magnitude below the ORCA sensitivity. However, for ORCA-100, there is a chance to identify it. The gamma-ray emission follows $t^{-1/3}$ as RS Oph.

\begin{figure}
    \centering
    \includegraphics[width=0.5\textwidth]{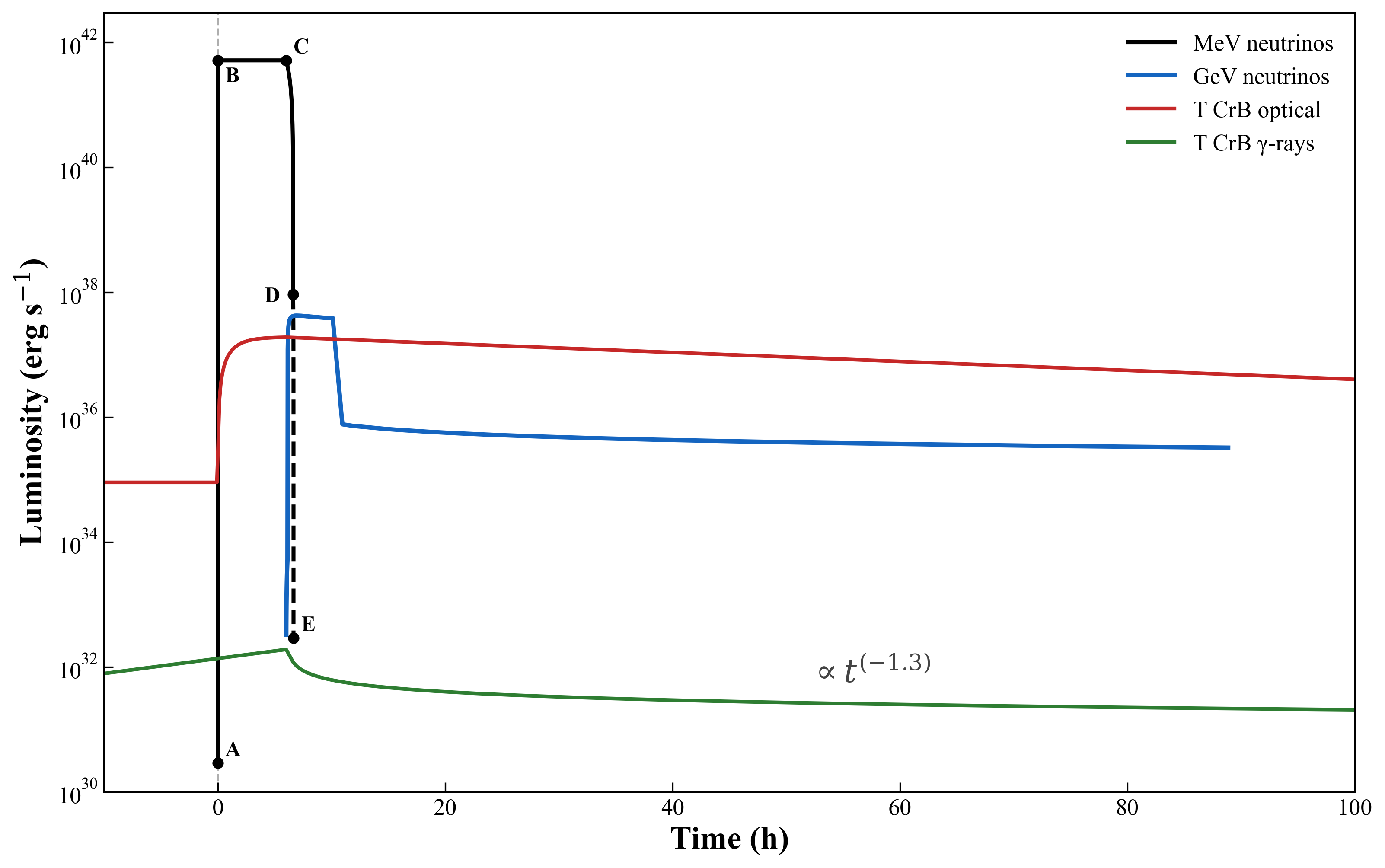}
    \caption{The predicted luminosity of T CrB for MeV (black) and GeV (black) neutrinos. Estimated optical (red) and gamma-ray (green) light curves for the next outburst of T CrB.}
    \label{T CrB Neutrinos Light-curve}
\end{figure}

\begin{table}[htbp]
    \centering
    \caption{The expected signal and background numbers of events in ORCA for RS Oph and T CrB at different phase.}
    \label{tab:the expected numbers of events}

    \begin{tabular}{cccc}
        \toprule
        Phase & RS Oph & T CrB & Background \\
        \midrule
        Phase1(\(T_1\)+0.2h) & $2.03 \times 10^{-3}$ & $5.72 \times 10^{-4}$ & $3.83 \times 10^{-9}$ \\
        Phase2(\(T_1\)+4h)   & $1.25 \times 10^{-1}$ & $3.51 \times 10^{-2}$ & $7.67 \times 10^{-8}$ \\
        Phase3(\(T_1\)+8h)   & $4.53 \times 10^{-3}$ & $1.13 \times 10^{-3}$ & $1.53 \times 10^{-7}$ \\
        Phase4(\(T_1\)+24h)  & $1.19 \times 10^{-2}$ & $1.81 \times 10^{-2}$ & $4.60 \times 10^{-7}$ \\
        \bottomrule
    \end{tabular}
\end{table}

Table~\ref{tab:the expected numbers of events} presents expected numbers of neutrino event for RS~Oph and T~CrB across different phases. The expected numbers of neutrino event is predominantly concentrated in the early stages of the outburst, peaking around Phase~2 with event rates reaching the order of $10^{-2}$. Furthermore, RS~Oph consistently exhibits slightly higher event rates than T~CrB.

\subsection{Implications for multi-messenger observations}
The multi-messenger observations of SymRNe can provide extremely useful information regarding the geometrical structure, shock evolution, and source environment. The MeV neutrinos emitted immediately during the TNR provide the alert to other observations and trace the thermonuclear burning on the WD surface. They take the dominant energy release during the first day. The rise time of the optics tells the thickness and density of the envelope. 

The strong dependence of the GeV neutrino flux on the density of the accretion disk suggests that high-energy neutrinos can serve as a probe of the immediate environment around the WD. In particular, an enhanced early-time neutrino signal would indicate the presence of a dense equatorial structure, whereas a weaker signal would be consistent with a more diffuse wind-dominated environment.

The close connection between hadronic gamma-ray and neutrino production implies that gamma-ray observations can be used as a trigger for neutrino searches. The detection of GeV gamma rays by instruments such as \textit{Fermi}-LAT can provide early evidence for particle acceleration, thereby motivating targeted searches for coincident neutrino emission. This coordinated approach is especially important given the short duration of the high-flux phase.

Finally, the results highlight the importance of rapid and coordinated synergy observations across different wavelengths and messengers. Since the most favorable conditions for neutrino detection occur shortly after the onset of the eruption, timely identification of nova events and prompt follow-up observations are essential. Future improvements in both detector sensitivity and alert systems will be crucial for fully exploiting the multi-messenger potential of SymRNe. The combined information will resolve the mystery of system structure and physical process. 

\section{Discussion and conclusion} \label{Discussion and limitations}
In this work, we construct a SymRN model including a dense and geometrically thin disk around the WD. It is motivated by the high accretion rates and by previous studies suggesting the existence of equatorial density enhancements. However, the distribution of the disk remains uncertain. If the disk is partially or fully disrupted, the enhancement of the GeV neutrino emission may be reduced. Conversely, a more extended or denser disk would lead to an even stronger early-time signal. In the absence of a disk, the ejecta expand into a more nearly spherical medium, yielding weaker and more isotropic neutrino emission. This distinct signature suggests that the GeV neutrino signal may serve as the sole unambiguous diagnostic for the presence of an accretion disk.

Moreover, the disk geometry is expected to depend on the orbital period of the WD binary. Longer orbital periods generally imply more substantial accretion \cite{Lechuga:2025jei}, leading to larger disks with wider opening angles and a more prominent equatorial density enhancement. The opening angle primarily redistributes the neutrino emission anisotropically: larger opening angles increase the observable flux for observers near the orbital plane while reducing it along polar directions. Therefore, for the next outburst of T CrB, upcoming observations of this system will offer an important opportunity to test the theoretical multi-messenger framework presented here. For example, if its GeV neutrinos are detected or the observed gamma rays are less than predicted, it means the disk should be larger and wider than predicted in our model. 

Moreover, we develop a time-dependent multi-energy model for neutrino emission from SymRNe, with RS Oph adopted as a representative system. 
{\color{black}The predicted gamma-ray spectrum, derived from the same proton distribution that produces the neutrinos with our model, is consistent with the observed GeV-TeV gamma-ray data from Fermi-LAT and H.E.S.S. during the 2021 outburst of RS Oph \cite{MAGIC:2022rmr,HESS:2022qap} as shown in Figure~\ref{fig:combined}.
Furthermore, the timescale analysis confirms that particle acceleration occurs efficiently before cooling dominates, and the interaction efficiency is sufficient to produce the predicted neutrino and gamma-ray fluxes. The energy budget and multi-wavelength constraints are all consistent, validating the physical consistency of our time-dependent multi-energy emission model.}
The neutrino emission exhibits a characteristic two-component temporal structure. The MeV neutrino signal rises rapidly during the TNR phase and directly traces the evolution of nuclear burning on the WD surface. For RS Oph, the companion optical emission is about 27.6 hours later than MeV neutrino, due to the scattering and diffusion. However they are absorbed only within the disk region after escape, leaving the majority largely unaffected.  The GeV neutrino emission appears with a delay and is governed by the interaction between the expanding ejecta and the surrounding CSM. While co-produced gamma rays near the WD are largely absorbed by the disk, consistent with the gamma-ray observations that are mainly from later time. 

Although current detectors have achieved great success \cite{IceCube, KM3NeT}, there is still a long way detecting SymRNe neutrinos. At the moment of TNR triggering, the nuclear interaction is very intense and a large number of MeV neutrinos emerge, but it is still not enough to be observed, even though the most advanced Super-Kamiokande neutrino detectors in this energy range are 4-5 orders of magnitude behind. 
The GeV component may become accessible for nearby systems with next-generation neutrino observatories such as KM3NeT/ORCA. The early-time phase of the eruption is identified as the most favorable window for detection, highlighting the importance of rapid observational follow-up. 
{\color{black}However, the predicted GeV neutrino flux remains below the IceCube sensitivity, which is consistent with the non-detection reported for T CrB \cite{IceCube:2025egb}. This underscores the necessity of next-generation detectors like ORCA\(\times\)100, which are optimized for the 1-100 GeV range.}
{\color{black} Furthermore, high-angular-resolution X-ray and radio observations offer a promising avenue to constrain the geometrical distribution of the nova environment, including the accretion disk, through their characteristic light-curves.\cite{Halzen:2019qkf}}

Overall, our results demonstrate that SymRNe are promising targets for multi-messenger astrophysics. The synergy of neutrino, optical, and gamma-ray observations can provide complementary information on particle acceleration, shock dynamics, and the structure of the circumstellar environment. Future improvements in detector sensitivity and coordinated observational strategies will be essential for fully exploiting this potential.

\begin{acknowledgments}

We gratefully acknowledge the contributions of the AAVSO observer community, whose photometric data and metadata resources were used in this study and made available through the AAVSO's International Database. This work is supported by the National Natural Science Foundation of China (NSFC) grants 12261141691. 

\end{acknowledgments}

\bibliography{apssamp}
The data that support the findings of this study are openly available in the AAVSO International Database (AID) at https://www.aavso.org/data/, as cited in Ref. \cite{aavso_database}.
\end{document}